\documentclass[useAMS,usenatbib]{mn2e}

\usepackage{graphicx}
\usepackage{times}
\usepackage{natbib}
\usepackage{color}

\newcommand{\apjs}{ApJS}

\newcommand{\apjl}{ApJL}

\newcommand{\aap}{A \& A}

\newcommand{\enzo}{\it{\small ENZO}}

\begin{document}
 
\title[Turbulent pressure in the ICM] {The turbulent  pressure support in galaxy clusters revisited}
\author[F. Vazza, M. Angelinelli, T.W.  Jones, D. Eckert ,Br\"{u}ggen M., Brunetti G.,   Gheller C. ]{F. Vazza$^{1,2,3}$\thanks{E-mail: franco.vazza2@unibo.it}, M. Angelinelli$^{1}$, T. Jones$^{4}$, D. Eckert$^{5}$, Br\"{u}ggen M.$^{2}$,G. Brunetti$^{3}$,  Gheller C.$^{6}$ \\
$^{1}$ Dipartimento di Fisica e Astronomia, Universit\'{a} di Bologna, Via Gobetti 92/3, 40121, Bologna, Italy\\ 
$^{2}$ Hamburger Sternwarte, University of Hamburg, Gojenbergsweg 112, 21029 Hamburg, Germany\\
$^{3}$Istituto di Radio Astronomia, INAF, Via Gobetti 101, 40121 Bologna, Italy\\
$^{4}$ University of Minnesota Twin Cities Minneapolis, MN, USA\\
$^{5}$ Max-Planck-Institut f\"{u}r extraterrestrische Physik, Giessenbachstrasse 1, 85748 Garching, Germany\\
$^{6}$ Swiss Plasma Center EPFL, SB SPC-TH, PPB 313, 1015 Lausanne, Switzerland}

\date{Received / Accepted}
\maketitle
\begin{abstract}

Due to their late formation in cosmic history, clusters of galaxies are not fully in hydrostatic equilibrium and the gravitational pull of their mass at a given radius is expected not to be entirely balanced by the thermal gas pressure. Turbulence may supply additional pressure, and recent (X-ray and SZ) hydrostatic mass reconstructions claim a pressure support of $\sim 5-15\%$ of the total pressure at $R_{\rm 200}$. In this work we show that, after carefully disentangling bulk from small-scale turbulent motions in high-resolution simulations of galaxy clusters,  we can constrain which fraction of the gas kinetic energy effectively provides pressure support in the  cluster's gravitational potential. While the ubiquitous presence of radial inflows in the cluster can lead to significant bias in the estimate of the non-thermal pressure support, we report that only a part of this energy effectively acts as a source of pressure, providing a support of the order of  $\sim 10\%$ of the total pressure at $R_{\rm 200}$.  

\end{abstract}

\label{firstpage} 

\begin{figure*}
\includegraphics[width=0.97\textwidth,height=0.26\textwidth]{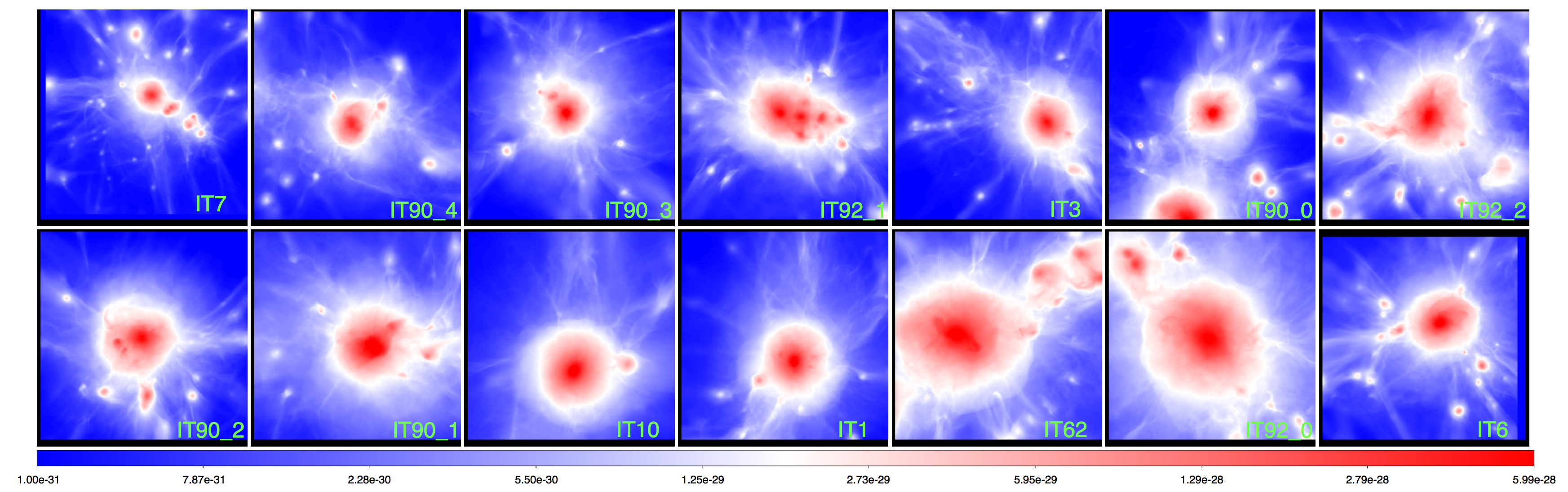}
\includegraphics[width=0.97\textwidth,height=0.26\textwidth]{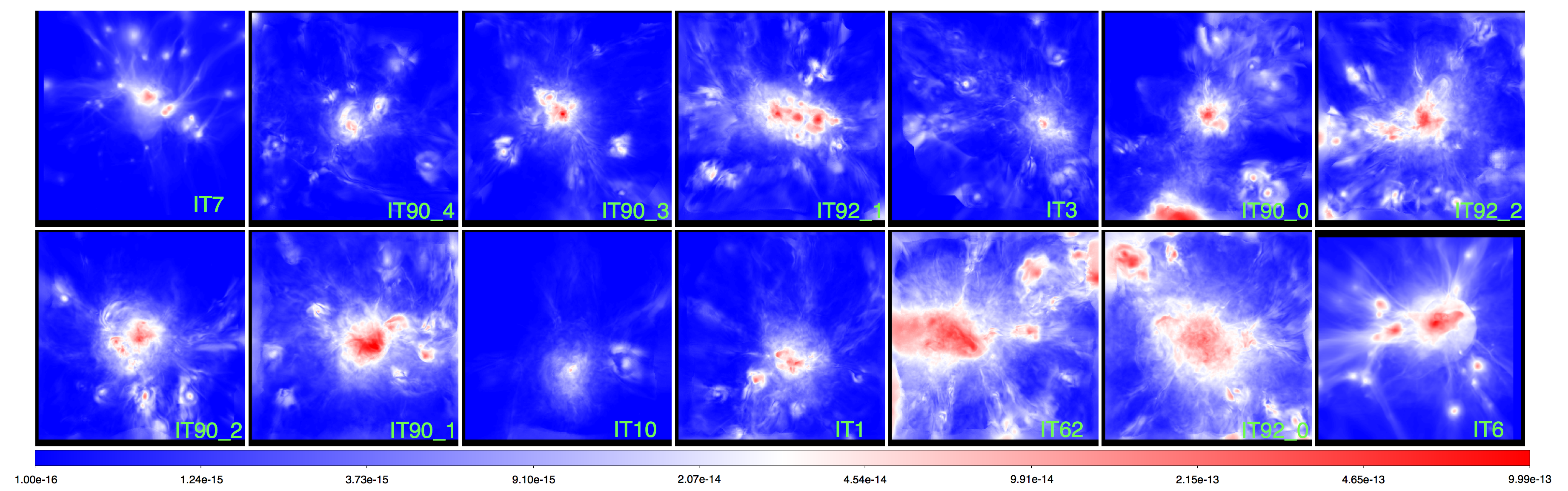}
\caption{Top two rows: mean projected gas density (in units of [$\rm g/cm^3$]) and (bottom two rows) mean projected turbulent kinetic pressure (in arbitrary code units) for each of our clusters at $z \approx 0$. Each image has a side of $4 \times 4 R^2_{\rm 100}$. The clusters are sorted in decreasing order (from top to bottom and from left to right) based on their $\langle w \rangle$ morphological parameter (see text).}  
 \label{fig:maps}
\end{figure*}

\section{Introduction}
\label{sec:intro}

The origin and evolution of turbulence  induced by the formation of large-scale structure has been studied with hydrodynamical simulations for more than a decade \citep[e.g.][]{do05,lau09,va11turbo,mi14,2014A&A...569A..67G}. Turbulence arises from the continuous stirring associated with
the growth of clusters, for example, via the injection and amplification of vorticity by shock waves 
\citep[e.g.][]{ry08,2015ApJ...810...93P,va17turbo} and
ram pressure stripping \citep[e.g.][]{su06,cassano05, 2007MNRAS.380.1399R}. 
Moreover, winds from star-burst galaxies, outflows from active galactic nuclei stir the intracluster medium (ICM), especially in cluster cores \citep[e.g.,][]{2005ApJ...628..153B,gaspari11a}.

However, direct measurements of turbulent gas motions in the ICM are rare. The Hitomi satellite managed to detect root-mean square velocities in the (fairly relaxed) Perseus cluster of $\sim 200 \rm ~km/s$ on $\leq 60 ~\rm kpc$ \citep[e.g.][]{hitomi,zuhone18}.
Highly resolved X-ray surface brightness fluctuations in clusters were interpreted as indications of moderate density fluctuations induced by the turbulent shaking of the ICM \citep[e.g.][]{sc04,2012MNRAS.421.1123C,2014A&A...569A..67G,2014Natur.515...85Z}. Moreover, hints of a correlation between X-ray surface brightness fluctuations and diffuse radio emission have recently been found \citep[][]{eck17,bo18}. This can be taken as evidence that the turbulence to which the X-ray surface brightness fluctuations bear testament powers the diffuse radio emission via turbulent re-acceleration \citep[e.g.][]{bl11b}.
Finally, the mass modeling of several galaxy clusters based on X-ray profiles suggested the presence of non-negligible  non-thermal pressure support potentially associated with ICM turbulence \citep[][]{2011MNRAS.416.2567M,Parrish2012,2018MNRAS.475.1340F,2018PASJ...70...51O}. 
Assessing the budget of turbulence in the ICM is key to correctly measure the mass of galaxy clusters.  For clusters that have not been disturbed by a recent merger, the ICM should be in hydrostatic balance, meaning that the gravitational pull of the gas is balanced by the total pressure gradient. The determination of density and temperature via X-ray observations can be used to measure the cluster's total gravitational mass.  To address this issue,  cosmological simulations have been used to constrain the level of turbulent pressure support that contributes to systematic errors in the hydrostatic mass estimates \citep[e.g.,][]{Kay04,Faltenbacher05, Rasia06,ha06,2007ApJ...668....1N}, which in turn may complicate the determination of cosmological parameters from galaxy clusters (e.g. $\sigma_8$ and $\Omega_{\rm M}$).

Recently,  \citet{eck18} have systematically analysed the hydrostatic mass bias in a sample of 14 galaxy clusters observed with the large XMM program X-COP \citep[][]{2017AN....338..293E}, providing evidence of an overall small level of non-thermal pressure support at $R_{\rm 200}$ and $R_{\rm 500}$ of order $\sim 5-15 \%$ of the total pressure. This non-thermal pressure contribution was found to be a factor $\sim 2-3$ below the 
expectations from most simulations \citep[e.g.][]{lau09,va11turbo,2014ApJ...792...25N,2016ApJ...827..112B}.  
 
In this paper we will revisit the measurement of 
non-thermal pressure produced by gas motions in the ICM.
Using recent high-resolution, Eulerian simulations of galaxy clusters, we show that the hydrostatic mass bias suggested by joint X-ray and SZ observations can be related to the fraction of the total gas kinetic energy that effectively act as a source of pressure support, after 
distinguishing cleanly between isotropic turbulent velocities and bulk motions.

The paper is structured as follows: in Sec.~\ref{methods} we describe our cluster sample and our recipes to isolate turbulent motions in the simulated ICM; in Sec.~\ref{sec:res} we give our results from the analysis of our sample and in Sec.~\ref{sec:conclusions} we discuss the limitations of our analysis and its implications for the interpretation of observations.

\section{Methods}
\subsection{The Itasca Simulated Cluster sample}
\label{methods}

We used the {\it "Itasca Simulated Clusters"} sample (ISC) for this project \footnote{http://cosmosimfrazza.myfreesites.net/isc-project. }, i.e.  a set of 14 galaxy clusters in the $5 \cdot 10^{13} \leq M_{\rm 100}/M_{\odot} \leq 6 \cdot 10^{14}$ mass range  simulated at uniformly high spatial resolution with Adaptive Mesh Refinement and the Piecewise Parabolic method in the {\enzo} \citep[][]{enzo14}.  Our simulations are non-radiative and assume the 
 WMAP7 $\Lambda$CDM cosmology \citep[][]{2011ApJS..192...18K}, with $\Omega_{\rm B} = 0.0445$, $\Omega_{\rm DM} =
0.2265$, $\Omega_{\Lambda} = 0.728$, Hubble parameter $h = 0.702$,  $\sigma_{8} = 0.8$ and a primordial index of $n=0.961$.  For each cluster, we generated two levels of nested grids as initial conditions (each with $400^3$cells and dark matter particles and covering $63^3 ~\rm Mpc^3$ and $31.5^3 \rm~Mpc^3$, respectively). At run time, we imposed 
two additional levels of {\it static} mesh refinement in the  $6.3^3 \rm ~Mpc^3$ subvolume around each cluster, down to  $\Delta x = 19.6 ~\rm kpc/cell$.  More information on the ISC sample are found in \citet{va17turbo} and \citet{wi17}. 

\begin{figure}
\includegraphics[width=0.495\textwidth,height=0.31\textwidth]{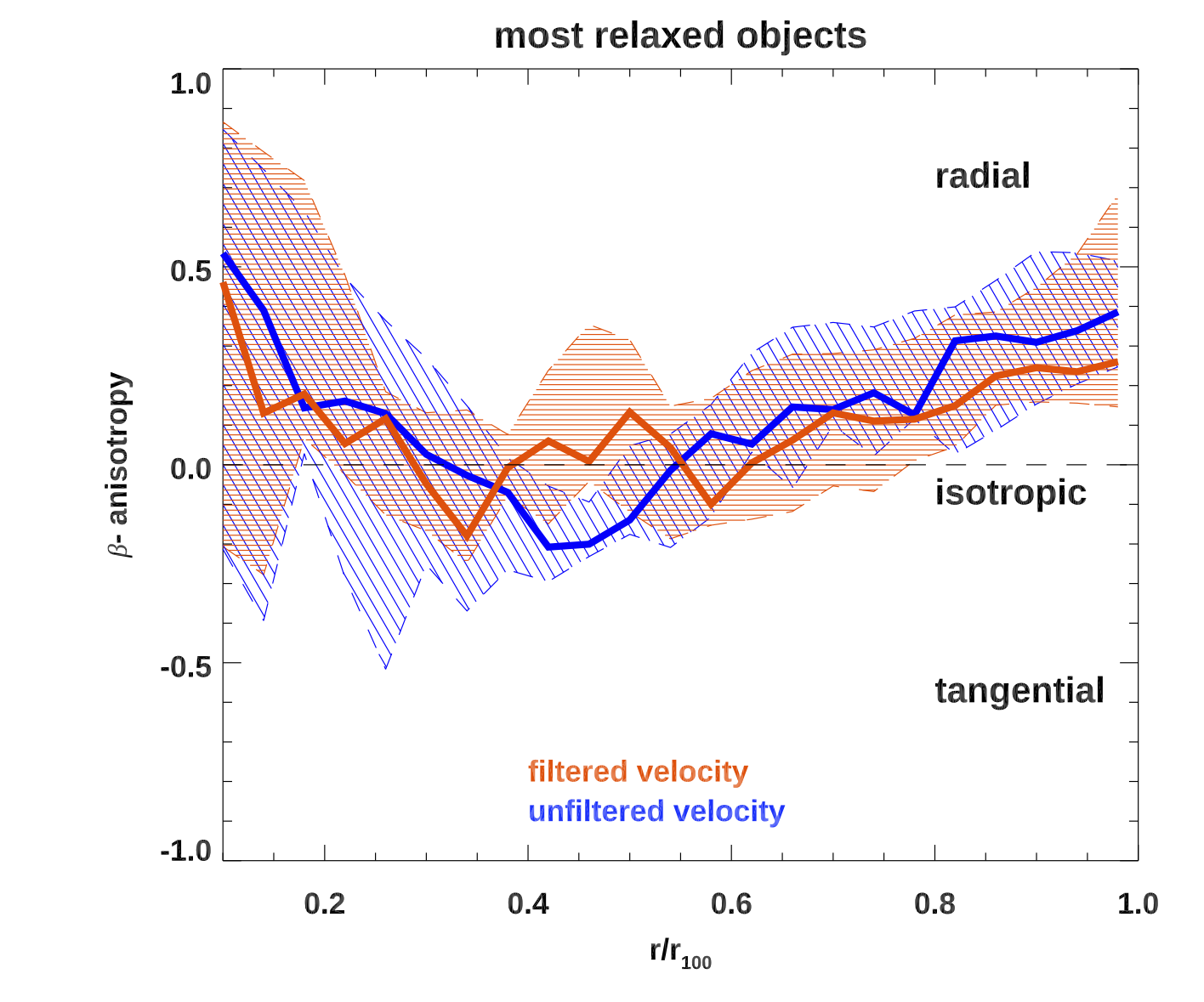}
\includegraphics[width=0.495\textwidth,height=0.31\textwidth]{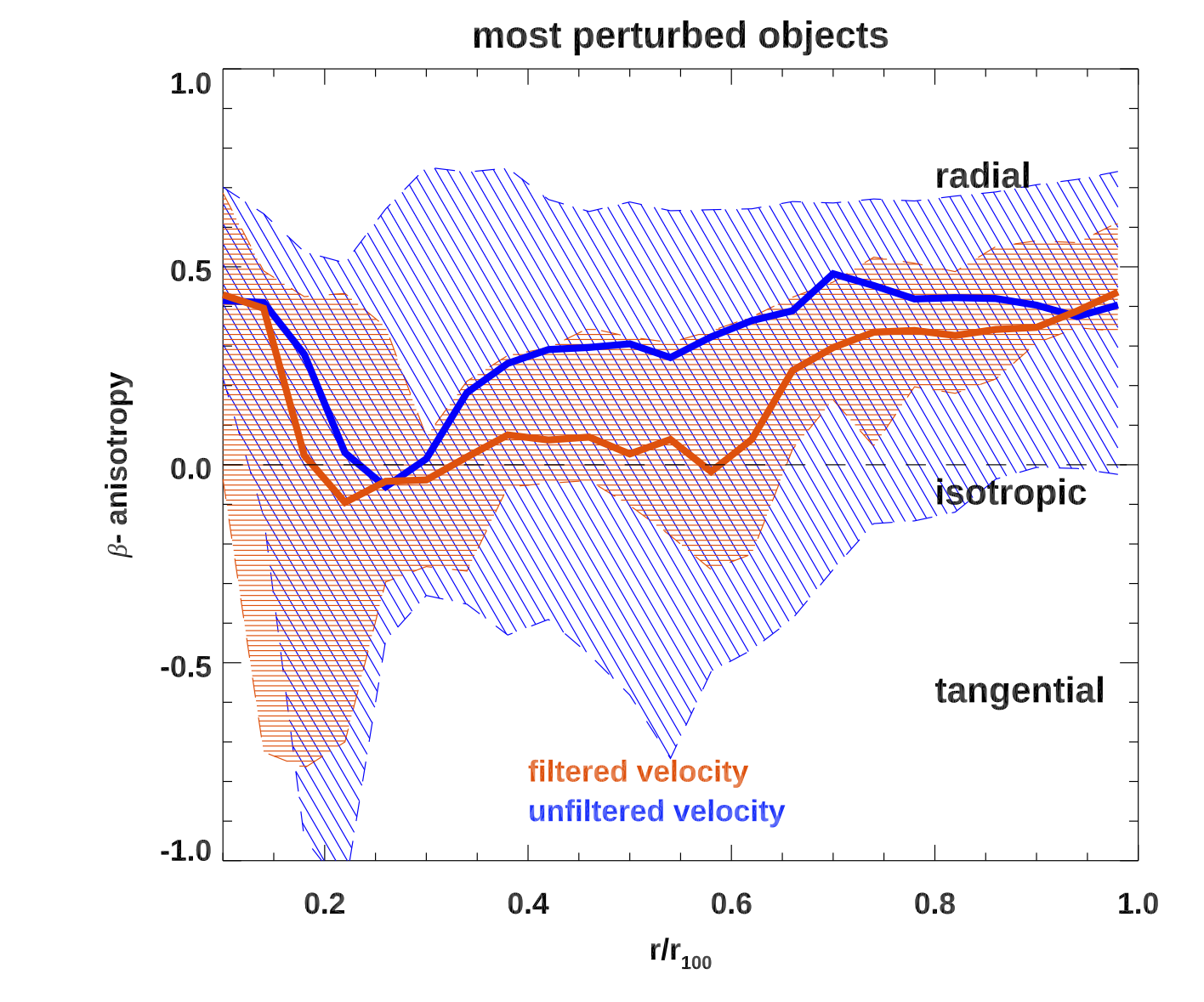}
\caption{Radial profiles of the anisotropy parameter, $\beta$, for a subset of our most relaxed and most perturbed clusters in the sample, with $25$ and $75$ percentiles shown with shadowed area.}  
 \label{fig:beta}
\end{figure}

\subsection{Identifying turbulence in the ICM}
\label{filter}

Several filtering techniques to identify turbulence in the complex ICM velocity fields have been developed over the years  \citep[e.g.][]{do05,in08,va12filter,mi14}. Here we rely on a combined set of
methods,  applied in post-processing, following \citet[][]{va17turbo}. Our main steps are: 

\begin{itemize}
\item {\it multi-scale filtering of turbulence}:  We applied the iterative,
multi-scale velocity filter from \citet{va12filter}, in which local mean (density weighted) velocity field for each cell, $\vec{V_L}$, is iteratively computed (separately for each velocity component) within  a domain of radius, $ L$. The small-scale residual velocity  fluctuations are computed as  $\vec{\delta v}=\vec{v}-\vec{V_L}$ for an increasing domain radius, until the relative change in $\vec{\delta v}$  between iterations falls below a  1\% tolerance.  The  iterations can also be stopped if  a shock stronger than our fiducial  ${\mathcal M_{\rm thr}}$ (see next item) enters the domain, as in \citet{va17turbo}.   The resulting $\vec{\delta v_L}$ gives our fiducial  estimate for the turbulent velocity magnitude for eddies of size $\approx 2 ~L$. Then the combination $\epsilon=(\delta v_L)^3/L \approx \epsilon_0$ estimates the dissipation rate of kinetic energy per unit mass, according to Kolmogorov theory (scale invariant by construction).  We remark that even if the stencil of cells used by our filter to constrain the local velocity field increases in an isotropic way, the algorithm can still detect anisotropic velocity structures, given its low tolerance ($1\%$), i.e. the filter does not bias the reconstructed small-scale fields to be isotropic (see Sec.~3). On the other hand, if steep velocity gradients are present, a fraction of the associated energy may be mis-identified as turbulent, as it mimics a velocity structure increasing as a function of scale.  However, based on the tests in \citet{va12filter} (Sec.~2.1), this small effect is expected not to be a relevant source of error, under realistic ICM conditions.
We note that in the forthcoming analysis, all quoted rms turbulent velocities must be referred to their specific scale, $L$, which is typically $\sim 200-400 ~\rm kpc$ for the range of masses analysed here, even if a distribution of turbulent scales is present in every cluster \citep[e.g.][]{va12filter,va17turbo}.

\item {\it shock identification}: Shocks are identified based on the 3D velocity jumps across cells . The  shock centre is given by the minimum in the 3D velocity divergence and the shock's Mach number  is constructed by combining the three velocity jumps from the Rankine-Hugoniot conditions in one dimension \citep[see][for more details]{va09shocks}. $\mathcal{M} \geq {\mathcal M_{\rm thr}}$ shocks are excised from our analysis (i.e. we avoid computing thermal and non-thermal pressure in such cells) in order to limit the contribution from velocity fluctuations related to shock-induced velocity fluctuations. We set ${\mathcal M_{\rm thr}}=3.0$,  higher than in our previous work (${\mathcal M_{\rm thr}=1.3}$) because here we focus on cluster outskirts, where the fraction of transonic  motions driven by accretion is larger than near the centre. 

\item {\it clump excision}:  Dense clumps associated with infalling structures can introduce a bias in the estimate of the local velocity field, as they correlate with large (and mostly laminar) bulk motions in the ICM. 
Observationally, clumps are generally masked when they are detectable in X-rays, and we follow a procedure for this similar to \citet{2013MNRAS.428.3274Z} and \citet{2013MNRAS.432.3030R}, masking the $10  \%$ densest cells (considering the gas density) at each radius from the cluster center. As for ${\mathcal M} \leq {\mathcal M_{\rm thr}}$ cells, we do not use these cells to compute the ratio of non-thermal to total gas pressure in our clusters,  which is also in line with what has been done in the X-ray analysis by \citet{eck18}, which serves as a comparison in the next Section.

\end{itemize}
In order to obtain the turbulent gas velocity, the above procedure is performed for each cluster at $z=0$. The projected turbulent pressure for all clusters in the ISC sample is shown in the lower panels of Fig.~\ref{fig:maps}.  The ratio between the turbulent pressure $P_{\rm NT}$ and the total pressure $P_{\rm tot}$ within each cell is thus:
\begin{equation}
\frac{P_{\rm NT}}{P_{\rm tot}} = \frac{\rho ~ \delta_v^2/\alpha_r}{(\rho ~ \delta_v^2 /\alpha_r+\rho k_{B} T/\mu_e m_p)}~, 
\label{pnt}
\end{equation}
where  $\rho$ the gas density, $\mu_e=0.59$ the mean molecular mass per electron and $T$ is the gas temperature. $\alpha_r$ is a numerical coefficient that can be either $\alpha_r=1$ if we only restrict to the radial velocity component  at each radius, or $\alpha_r=3$ if we simply assume an isotropic velocity dispersion at each radius. 
The resulting profile  of non-thermal pressure ratio is thus: 
\begin{equation}
X(R) = \frac{\sum_i P_{\rm NT,i}}{\sum_i P_{\rm tot,i}}~,
\label{prof}
\end{equation}
in which the summation refers to all cells within each radial shell after the excision of shocks and clumps.

 In the following, we will refer to the small-scale filtered, clump excised and shock masked velocity field as to the "turbulent" velocity , and to the clump-excised only velocity field as to the "unfiltered" velocity.

\begin{figure*}
\includegraphics[width=0.995\textwidth,height=0.125\textwidth]{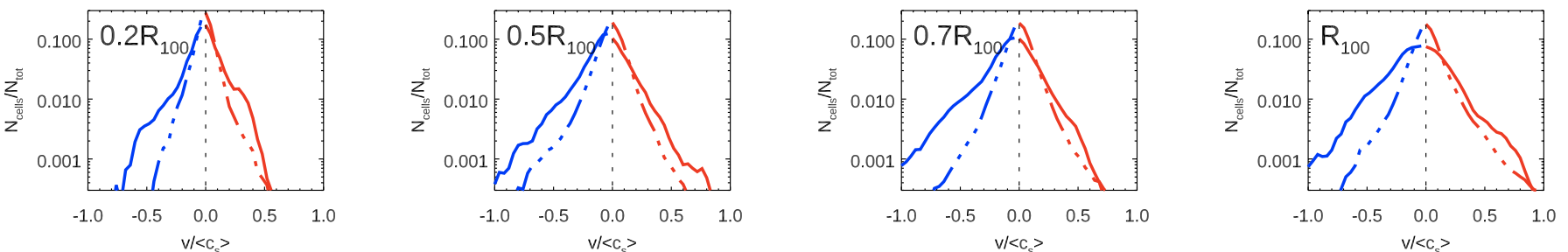}
\caption{Distribution of velocities at different radii in our cluster sample, normalized to the average sound speed within each shell. The solid lines give the distribution for the total (unfiltered) velocities, the dot-dashed line give the distribution for the turbulent (small-scale filtered) velocities.}  
 \label{fig:PDF2}
\end{figure*}

\section{Results}
\label{sec:res}

Previous ICM simulations have proven that the turbulent pressure support from gas motions is increasingly anisotropic (and dominated by radial motions) away from the cluster centers \citet{lau09}. It has also been shown that that the pressure support from  large-scale rotational motions is generally small in clusters \citep[][]{Fang2009,2013ApJ...767...79S,2013ApJ...777..151L,2014ApJ...792...25N}. This stems from the fact that the pressure scale height  is in general smaller than the cluster core radius ($H_p=(d ln P/dr)^{-1} \sim 50-200 ~\rm kpc$ for the masses explored here), meaning that outside of cluster cores regular gas motions along the radius are  dissipated into smaller-scale turbulent motions, as directly measured in the simulated ICM. 

The level of anisotropy of gas motions is usually characterized through the anisotropic parameter $\beta=1-\sigma^2_t/2\sigma^2_r$ ($\sigma_r$ and $\sigma_t$ are the gas velocity dispersion in the radial and tangential direction to the cluster centre, respectively), hence in general a velocity dispersion in the radial direction is representative of the true local turbulent pressure only if $\beta \sim 0$. 

Fig.~\ref{fig:beta} shows the (density-weighted) profiles of $\beta(r)$ for the five most perturbed and the five most relaxed clusters in our sample, with ranking based on their  morphological parameter $\langle w \rangle$ (averaged over the three lines of sight). This parameter quantifies the shift of the X-ray centroid as a function of cluster radius
\citep[e.g.][]{1993ApJ...413..492M}, which is here measured on bolometric X-ray maps for simplicity. In this analysis, we rely on the $\langle w \rangle$ parameter instead of other morphological parameter such as the concentration parameter, $c$, or the power ratio, $P_3/P_0$, as we know that $w$ is the parameter that best correlates with the non-thermal pressure at large cluster radii (Angelinelli et al, in preparation). 
Fig.~\ref{fig:beta} shows that our filtering approach reduces the scatter in $\beta$ significantly, in particular for perturbed clusters. The filtering of velocity also reduces the anisotropy  in the $0.2-0.9 ~\rm R_{\rm 200}$ range compared to the unfiltered velocity field, even if an excess of radial bias even at small scale remain in the outermost accretion regions.

At certain fixed radii ($0.2$, $0.5$, $0.7$ and $1.0$ times the $R_{\rm 100}$ of each cluster) we measured the distribution of total (i.e. unfiltered) and filtered radial (centred on the cluster) velocities. The excision of the densest $10\%$ cells within each radial shell is performed on both total and turbulent velocities to remove the effect of single dense clumps.  Fig.~\ref{fig:PDF2} shows the distribution of the radial velocity components averaged across all clusters, where we normalized the velocity to the average sound speed within each radial shell, measured within $R_{\rm 100}$, in order to take account for the different masses across the sample. The unfiltered radial velocity field systematically displays a large degree of asymmetry towards the "blue-shifted" part of the distribution ($v_r < 0$), meaning that in most clusters there is a preference for radial motions pointing towards the cluster center,  which is also confirmed by the fact that we measure $dv^2/dr > 0 $ is  at most radii in our clusters (not shown). The effective kinetic pressure component that is caused by laminar inflows clearly reduces the pressure support over what is needed for hydrostatic equilibrium.  This kinetic pressure acts in the direction of the gravitational force of the host cluster, and  it effectively pushes gas {\it inwards}, opposite to an isotropic pressure component. 
However, the small-scale filtered velocity displays a  more marked symmetry at each radius, indicating that the velocity fields extracted in such a way are indeed fairly symmetric in the radial direction and thus act as a true non-thermal pressure component, at least on the $\sim 200-400 ~\rm kpc$ scales reconstructed by our analysis. 

At the radii that are presently best probed by X-ray observations, e.g. $R_{\rm 200} \approx 0.7 R_{\rm 100}$ and $R_{\rm 500} \approx 0.5 R_{\rm 100}$  \citep[e.g.][]{eck18}, the "blue-shifted", inward component of the velocity field shows an excess of the order of a factor $\sim 2$ compared to the symmetric small-scale filtered component.  Even larger discrepancies between the filtered and non-filtered distribution of velocities are found in specific objects, with an increased departure from a Gaussian distribution of radial velocity components in perturbed objects. 
If we incorrectly assume instead that rms of the unfiltered velocity at each radius stems from a symmetric Gaussian distribution, then the associated $X(R)$ would be overestimated.  This could explain the systematic overestimate of the non-thermal pressure support reported by most cosmological simulations to date. 

\begin{figure}
\includegraphics[width=0.499\textwidth,height=0.34\textwidth]{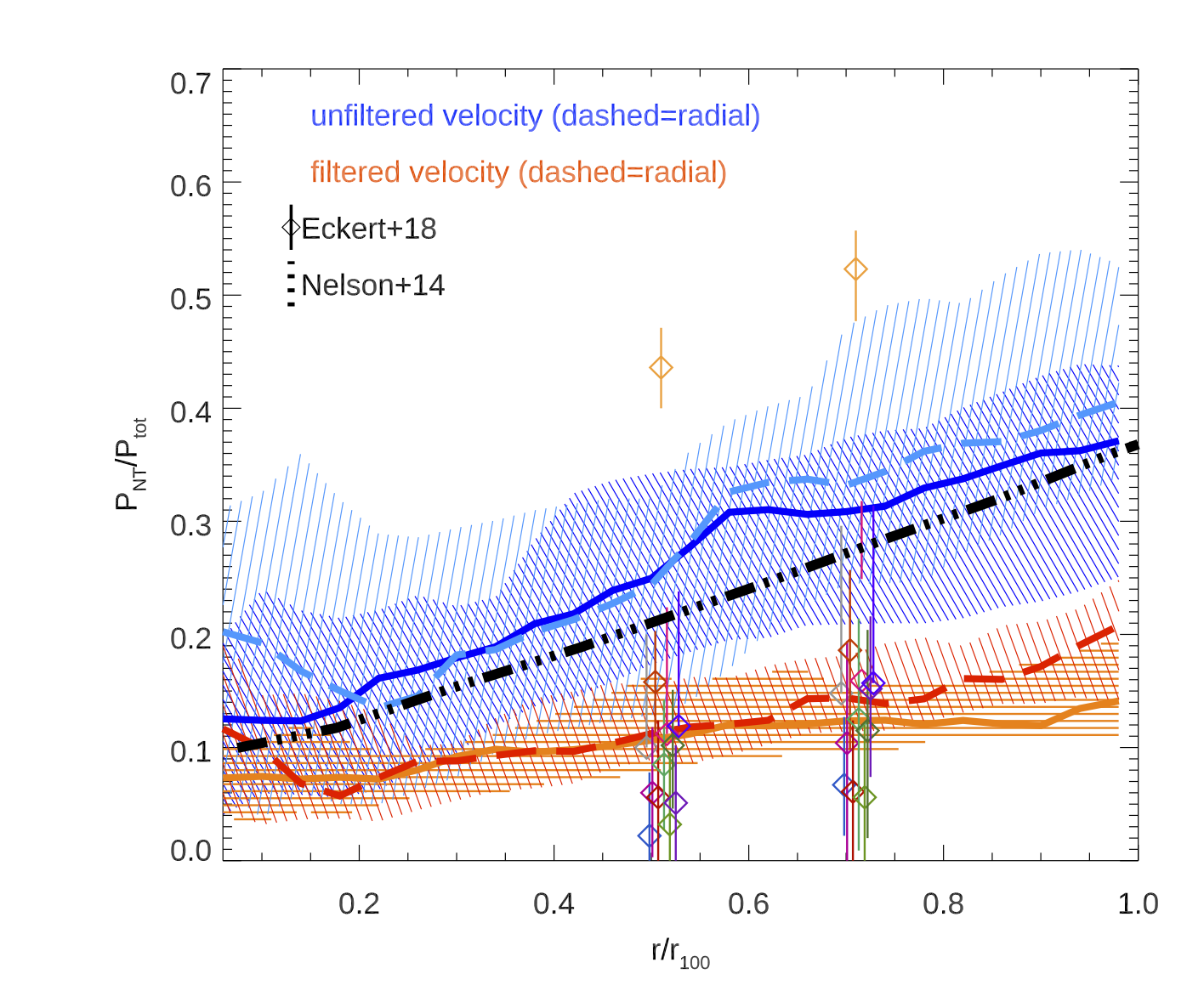}

\caption{Radial profiles of $X(R)$  for our sample at $z=0$, for the full kinetic energy (blue) and the filtered turbulent energy (orange). The solid lines assume an isotropic velocity distribution along the radius, while the dashed lines only consider the truly radial component. The shadowed areas give the $25-75$ percentiles around the median. We also show  the data from \citet[][]{eck18} observations, and from \citet[][]{2014ApJ...792...25N} simulations.}  
 \label{fig:prof}
\end{figure}

This conclusion is confirmed by Fig.~\ref{fig:prof}, which shows the central result of this work: there we present the average profile of the $X(R)$ ratio for the entire sample and for the filtered or unfiltered velocities. We plot, both, the radial pressure support from the rms velocity values at each radii, assuming isotropy ($\alpha_r=3$ in Eq.\ref{pnt}) as well as only considering the radial velocity component ($\alpha_r=1$). 
These results are contrasted with the recent observational estimates by \citet[][]{eck18} at $\approx R_{500}$ and $\approx R_{200}$ (symbols). For comparison with previous numerical work, we also show the best-fit profile for the non-thermal pressure from turbulent motions suggested by \citet{2014ApJ...792...25N}. 
The  profile of $P_{\rm NT}/P_{\rm tot}$ for turbulent velocities is much flatter compared to the unfiltered case, and falls within the the $\sim 5-15 \%$ level hinted by observations. With the exception of one observed system that clearly stands out of the rest of the distribution (A2319, \citealt{2018A&A...614A...7G}) and within the fairly limited statistics of the two sample (neither of which is a mass complete one), the observed and simulated estimates of non-thermal pressure support are in the same range.  A steeper trend with radius, as well as a $\sim 2-3$ times higher non-thermal pressure support would be instead inferred using the more standard unfiltered velocity field. While this work indeed confirms that this is the typical level of gas kinetic energy at this radius in average clusters, only $\sim 1/2-1/3$ of this energy is associated to isotropic and volume filling motions and acts as a source of pressure, and would thus be deduced from hydrostatic mass reconstructions.

\begin{figure}
\includegraphics[width=0.499\textwidth]{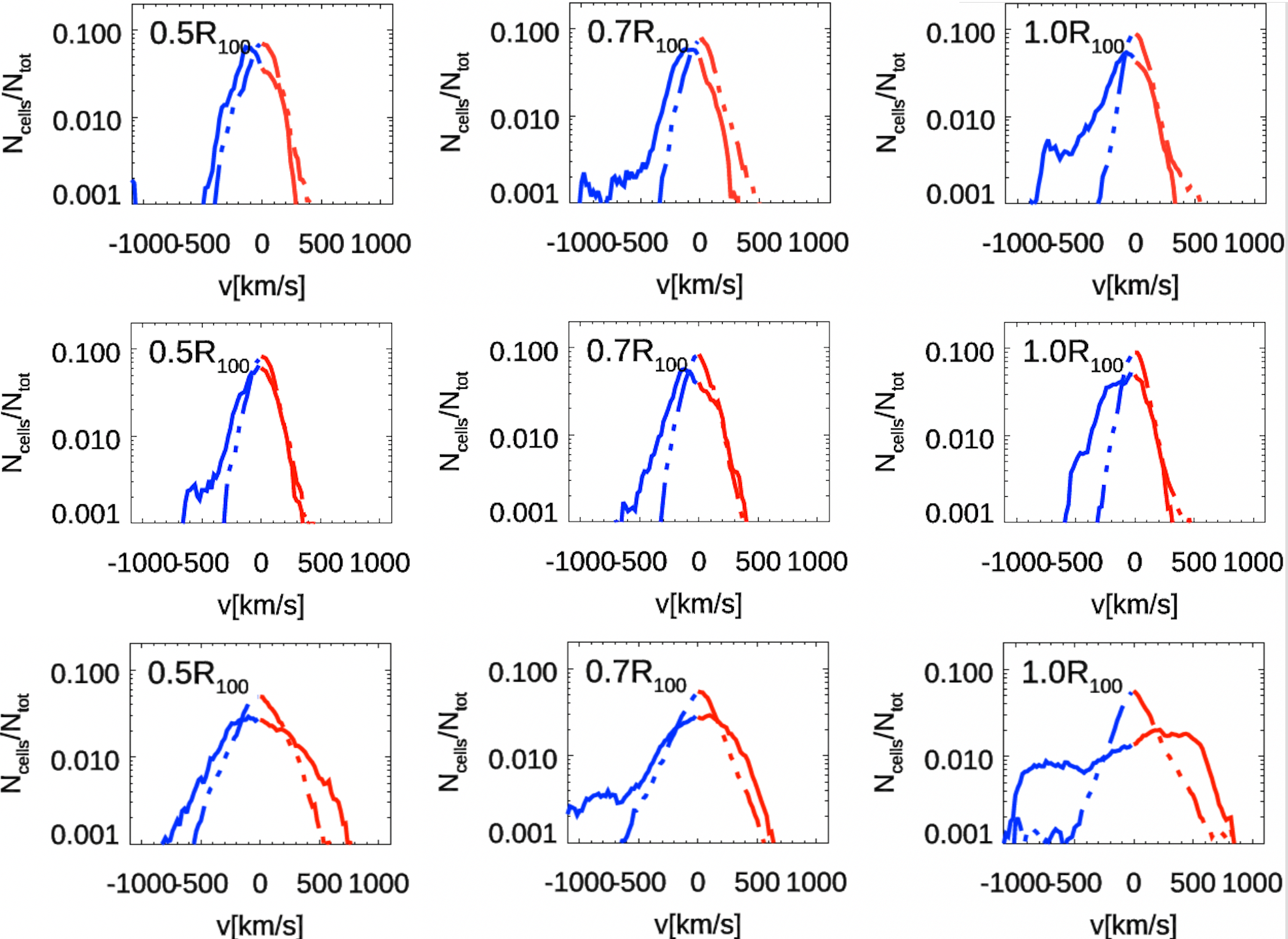}
\caption{Velocity distributions at different radial locations for cluster  IT90\_3 (merging), IT90\_4 (relaxed) and IT62 (post-merger).}  
 \label{fig:PDF3}
\end{figure}

\section{Discussion and Conclusions}
\label{sec:conclusions}

We investigated the distribution of kinetic (turbulent) pressure in the ICM, using a sample of recent high resolution simulations of (non-radiative) galaxy clusters \citep[][]{va17turbo,wi17}. 
In particular, motivated by recent measurements on the hydrostatic mass bias of XMM-Newton analysis \citep[][]{eck18}, we quantified the kinetic pressure support by residual gas motions in the ICM.
When properly analysed, the turbulent kinetic energy of the ICM is a small fraction ($\sim 1/2-1/3$) of the total kinetic gas energy at large radii, unlike what is usually estimated with more standard analysis. The effective pressure support from turbulence, after removing bulk motions, is on average $\sim 10\%$ of the total gas pressure. Although the presence of bulk motions that we detect and subtract in simulations may affect the estimate from X-ray observations in real clusters, we note that this is of the same order of what recently suggested by joint X-ray and SZ observations \citep[][]{eck18}. If this scenario is confirmed,  no additional mechanism to the standard modelling of the ICM on $\geq 20 ~\rm kpc$ (e.g. increased viscosity or enhanced turbulent dissipation)  appears necessary in order to reconcile with the most recent hydrostatic mass reconstructions. 
  
\section*{acknowledgements}

We wish to dedicate this work to the memory of Giuseppe (Bepi) Tormen, who tragically passed away on June 2018, and first introduced F.V. to cosmological simulations and the beautiful mess they contain.  We used the {\enzo} code (http://enzo-project.org), the product of a collaborative effort of scientists at many universities and national laboratories. We thank our reviewer, E. Churazov, for useful comments which improved the quality of our work, and A. Bonafede for performing the calculation of X-ray morphological parameters.  F.V. acknowledges financial support from the Horizon 2020 program under the ERC Starting Grant "MAGCOW", no. 714196. T.W.J. acknowledges support from the US National Science Foundation. The simulations were carried out at the Minnesota Supercomputing Institute at the University of Minnesota. 

\bibliographystyle{mnras}
\bibliography{franco}

\end{document}